# Partition-based K-space Synthesis for Multi-contrast Parallel Imaging

Yuxia Huang, Zhonghui Wu, Xiaoling Xu, Minghui Zhang,
Shanshan Wang, *Senior Member, IEEE,* Qiegen Liu, *Senior Member, IEEE*

*Abstract*—Multi-contrast magnetic resonance imaging is a significant and essential medical imaging technique. However, multi-contrast imaging has longer acquisition time and is easy to cause motion artifacts. In particular, the acquisition time for a T2-weighted image is prolonged due to its longer repetition time (TR). On the contrary, T1-weighted image has a shorter TR. Therefore, utilizing complementary information across T1 and T2-weighted image is a way to decrease the overall imaging time. Previous T1-assisted T2 reconstruction methods have mostly focused on image domain using whole-based image fusion approaches. The image domain reconstruction method has the defects of high computational complexity and limited flexibility. To address this issue, we propose a novel multi-contrast imaging method called partition-based k-space synthesis (PKS) which can achieve super reconstruction quality of T2-weighted image by feature fusion. Concretely, we first decompose fully-sampled T1 k-space data and under-sampled T2 k-space data into two sub-data, separately. Then two new objects are constructed by combining the two sub-T1/T2 data. After that, the two new objects as the whole data to realize the reconstruction of T2-weighted image. Finally, the objective T2 is synthesized by extracting the sub-T2 data of each part. Experimental results showed that our combined technique can achieve comparable or better results than using traditional k-space parallel imaging (SAKE) that processes each contrast independently.

## I. Introduction

Magnetic Resonance Imaging (MRI) is extensively employed in the clinic to diagnose various diseases and injuries and offers advantages such as non-ionizing radiation and soft tissue resolution [1, 2].[1] However, in clinical applications, it is often necessary to obtain images of the same anatomical structure with different contrasts, such as Proton Density-weighted image (PDWI), T1-weighted image and T2-weighted image (i.e., T1WI and T2WI) [3, 4]. Multi-contrast MRI can provide more structural and functional information, enabling accurate diagnosis and evaluation of different diseases. Each contrast image is used to examine the different physical properties of the same tissues [5, 6]. The T1WI highlights the tissue structure, which has a shorter TR. The T2WI prioritizes capturing fine details, but this comes at the expense of a slower image acquisition speed. Therefore, we can consider performing k-space under-sampling [7, 8] to obtain faster T2WI reconstruction.

Parallel imaging (PI) is based on the simultaneous acquisition of signals from multiple receiver coils to speed up scanning. Traditional PI reconstruction methods can be broadly categorized into two types. One in which the reconstruction method is performed in image space and the other type is performed in k-space. An image domain reconstruction example is sensitivity encoding (SENSE) [9], where SENSE required the estimation of the sensitivity distribution pattern of each receiver coil, and used coil sensitivity contours to recover the image from under-sampled data. Moreover, there are various k-space domain reconstruction methods available, such as generalized self-calibrating PI acquisition (GRAPPA) [10], iterative self-consistent PI reconstruction (SPIRiT) [11], efficient eigenvector-based SPIRiT (ESPIRiT) [12], simultaneous auto-calibration and k-space estimation (SAKE) [13], and PI low-rank matrix modeling in local k-space neighborhoods (P-LORAKS) [14].

In PI, both GRAPPA [10] and ESPIRiT [12] methods employ calibration data to estimate and restore missing k-space information. Unlike the above two imaging methods that require additional auto-calibration signal (ACS) lines for estimating reconstruction kernels or sensitivity maps, SAKE [13] avoided this step by integrating auto-calibration and k-space estimation within a unified framework. It employs a projection-set algorithm with singular value thresholding to recover the missing k-space data by alternately enforcing data consistency and low-rank Hankel matrix structure without any fully-sampled calibration region. While PI methods improve image quality and reduce data acquisition time, most of the methods reported in the literature focus on reconstructing MR images using under-sampled data from the same contrast. There has been a lack of emphasis on investigating the relationship between different MR sequences and fully leveraging the redundant information available in multi-contrast datasets [15-18]. Therefore, for the problem of T1/T2WI, we show that further exploiting the complementarity between different T1/T2WI can reconstruct more accurate images.

Leveraging the high-quality images obtained from the auxiliary mode (T1WI) to assist the target mode (T2WI) is essentially an accelerated imaging process, which has been confirmed by previous work. For example, Kim *et al.* [19] generated MR images by combining anatomical information from different contrast images, which improved the reconstruction performance. Alkan *et al.* [20] proposed a 9-layer two-dimensional convolutional neural network using T1WI and eight tissue masks as inputs to generate the corresponding T2WI. Vemulapalli *et al.* [21] solved the problem of cross-modal medical image synthesis in an unsupervised environment, generating T1-MRI scans from T2-MRI scans and vice versa. A similar work in [22], reconstructed high-quality T2WI based on the Dense-Unet deep learning method combining fully-sampled T1WI and under-sampled T2WI. However, these multi-contrast deep learning methods

---

This work was supported in part by National Natural Science Foundation of China under 62122033 and Key Research and Development Program of Jiangxi Province under 20212BBE53001. (Y. Huang and Z. Wu are co-first authors.) (Corresponding authors: S. Wang and Q. Liu)

Y. Huang, Z. Wu, X. Xu, M. Zhang, Q. Liu are with School of Information Engineering, Nanchang University, Nanchang 330031, China. ({huangyuxia, wuzhonghui}@email.ncu.edu.cn, {xuxiaoling, zhangminghui, liuqiegen}@ncu.edu.cn)

S. Wang is with Paul C. Lauterbur Research Center for Biomedical Imaging, SIAT, Chinese Academy of Sciences, Shenzhen 518055, China. (sophiasswang@hotmail.com)

are based on the image domain and image synthesis is a whole-based process [23]. In other words, the multi-contrast images are integrated, which do not reflect the partition concept and cause intensive computational workload.

To alleviate these deficiencies, we further explore this aspect and propose a novel multi-contrast method that incorporates the concept of partition. It is termed as **Partition-based K-space Synthesis (PKS)** for multi-contrast reconstruction with PI (see **Section III** for details). Our method combines with traditional PI technology SAKE [14] to apply multi-contrast MRI reconstruction, and performs partition processing on k-space data. It transforms the recovery problem of one data into the recovery problem of multiple data. To be specific, we generate multiple sub-data by segmenting the fully-sampled T1 k-space data and under-sampled T2 k-space data, respectively. Subsequently, we combine the sub-T1 data with the sub-T2 data to formulate new objects. After reconstruction, we can obtain complete T2 data by synthesizing sub-T2 data in each object. This enables us to effectively reconstruct T2WI with improved accuracy and resolution. Finally, experiments conducted on a multi-contrast dataset demonstrated that our proposed PKS multi-contrast model provided superior image quality compared to the single-contrast MRI algorithm. Our main **contributions** can be summarized as follows:

- *Super Partition-based Scheme:* Different from the previous multi-contrast methods which T1-assisted T2 imaging is a whole-based image fusion process, we introduce a concept of partition for the first time in the field of MRI. The PKS scheme includes various partition mechanism, the number of partitions such as partition-2 (partition into two equal parts by row or column), partition-3 and partition-4 (see Section III) that all achieve accurate reconstruction quality of T2WI.
- *Robust K-space Synthesis Model:* Different from the existing multi-contrast MRI methods in the image domain, we introduce a new k-space synthesis model, which uses a variety of contrasts (T1/T2) to synthesize target images. Our method effectively enhances fusion quality and strengthens the robustness of the model by using fully-sampled T1WI and under-sampled T2WI in k-space.

The remainder of this paper is organized as follows. Section II briefly overviews some preliminary works in this study. Section III contains the key idea of the synthesis approaches. The experimental settings and results are shown in Section IV. Section V conducts a concise discussion and Section VI draws a conclusion for this work.

## II. PRELIMINARY WORK

### A. Target Function Acquisition

Since our proposed model is processed in k-space, i.e., T2WI is reconstructed in k-space using fully-sampled T1 and under-sampled T2. Therefore, we need to obtain the objective under-sampled T2 k-space data. We propose under-sampled T2 k-space data as follows:

$$X_{T2} = M \odot Fy_{T2} \quad (1)$$

where $M$, $F$ and $y_{T2}$ represent the under-sampled mask, Fourier transform and acquired T2WI, respectively. $\odot$, $\otimes$ represent matrix point multiplication and numerical cross product operation. Subsequently, we can obtain the T2 zero-filled reconstruction image by performing the inverse Fourier transform of T2 k-space data.

$$x_{T2} = F^H \otimes X_{T2} \quad (2)$$

The segmented source data is obtained through a concatenation operation [24].

$$X_{T2}(syn\_source) = Concat \begin{pmatrix} X_{T2}(sub\_data1) \\ X_{T2}(sub\_data2) \end{pmatrix} \quad (3)$$

where sub_data1 and sub_data2 denote the upper/low half of the T2 data in the row-based partition mechanism, respectively. More detail introduction sees **Section III.**

### B. Traditional K-space based PI Reconstruction

SAKE [13] is an algorithm for reconstructing images from randomly under-sampled data without calibration data. Only uses all acquired data for calibration without separate calibration data. It does this by constructing the entire k-space data into a low-rank matrix of large Hankel structures and then recovering images from under-sampled multi-coil data. Specifically, it strings up the multi-coil data and re-represents the connected data as the Hankel matrix form for Cadzow's signal enhancement [25]. This process can be described as follows:

$$H_k = H(Y) \quad (4)$$

where $H_k$ represents a structured low-rank Hankel matrix, $H$ is a linear operator that generates a data matrix $H_k$ from multi-coil dataset $Y$ concatenated in a vector form. The resulting structured matrix has the low-rank (LR) property [26-28]. Therefore, the problem of reconstruction for PI, without the fully-sampled calibration regions, can be regarded as a low-rank matrix completion problem in k-space. If the known rank of the data matrix is $k$ (low-rank projection), it can be expressed as:

$$\min_X \|DX - Y\|^2 \quad (5)$$

$$s.t. \ rank(H_k) = k, \ X = H^\dagger(H_k) \quad (6)$$

where $Y$ and $X$ represent the k-space data acquired from the coil and the reconstructed k-space data to be derived, respectively, and then converted into image space. $D$ denotes the sampling operation related to the sampling matrix. $H$ is a linear operator that projects $X$ into the data matrix $H_k$ using a block Hankel matrix structure. $H^\dagger$ is the corresponding pseudo-inverse operator. In the iteration process, the data matrix passes through the LR projection, the structural consistency (SC) projection, and the data consistency (DC) projection in turn. The iteration continues until the number of repetitions reach the maximum or updates the reconstructed data within tolerance.

### C. Multi-contrast MR Reconstruction Methods

In the process of clinical application, multi-contrast imaging technology has become an important tool in the field of medical image diagnosis, which can provide doctors with different contrast images to help them make more accurate diagnosis, such as PDWI, T1WI and T2WI. In recent years, multi-contrast synthesis has attracted a lot of attention in MRI. At present, there are two main kinds of multi-contrast methods. One kind is CNN-based method [29, 30] and the other kind is GAN-based method [31-33]. The under-sampled image can be reconstructed with another contrast image as auxiliary information. For example, Murugesan *et al.* [34] improved structural recovery on T2WI with the help of T1WI. Yang *et al.* [35] used T1WI to assist T2WI reconstruction using conditional GAN. Kawahara *et al.* [36] used

GAN to establish a prediction framework for T1WI to T2WI and T2WI to T1WI. Furthermore, many studies have shown the benefits of assisted reconstruction. The reconstruction of T2WI is achieved by taking the fully-sampled T1WI and the under-sampled T2WI as input to the reconstruction network [37-38]. However, these two deep learning-based approaches require extensive training and testing time. In the case of multi-contrast synthesis, the process takes even longer. Moreover, the image domain fusion suffers from high computational complexity and lacks flexibility.

### III. METHOD

#### A. Overview of the Proposed Method

By acquiring different multi-contrast images from the same anatomical structure, we can obtain more diagnostic information. Most previous T1-assisted T2 reconstruction methods have concentrated on utilizing whole-based image fusion approaches within the image domain. However, these image domain-based reconstruction methods have some limitations. Firstly, they have relatively high computational complexity and require a lot of computational resources and time to complete the image fusion process. This can be a challenge for clinical applications with high requirements. In addition, the image domain-based reconstruction method has some difficulties in dealing with the nonlinear relationship between different contrast images, so it may not be able to make full use of the information of different images in some cases.

Therefore, to alleviate the ill-posedness of this problem, we divide this problem into two sub-problems. First, we process data in k-space, which not only avoids the nonlinear insufficiency between different contrast images when processing in the image domain, but also has greater flexibility. Second, we use the fusion of partition-based images instead of the fusion of the whole-based images, transforming the recovery problem of one data into the recovery problem of multiple data. This allows us to make full use of the complementary information between images with different contrasts. Fig. 1 shows the overall procedure of the proposed method. The PKS method can be divided into the following three steps. Specifically, in Step 1 denotes as PKS-Transform (PKS-T), we decompose under-sampled T2 k-space data into two parts and then combine them with sub-T1 data of the same size to form two new objects. In Step 2 denotes as PKS-Process (PKS-P), the two new objects are reconstructed separately using the SAKE algorithm. In Step 3 denotes as PKS-Inverse Transform (PKS-IT), we will carry out T2 extraction of the two reconstructed k-space data (Composed of the orange block module) to synthesize the target T2 value. Overall, it is of great significance to comprehensively apply these two contrasts according to the specific situation. Specific implementation details are introduced in **Section III. B**.

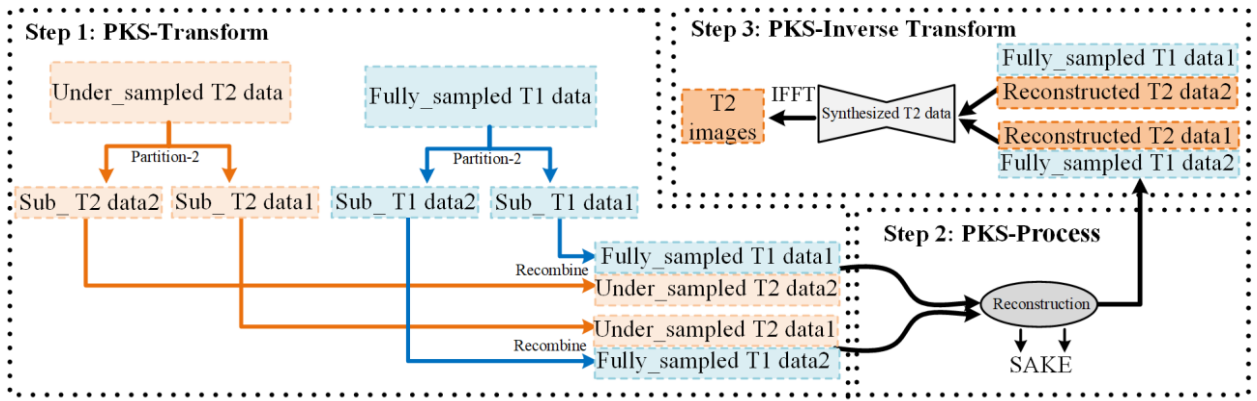

**Fig. 1.** Overview of the proposed method. The PKS method mainly consists of three steps. First, the under-sampled T2 data and fully-sampled T1 data are decomposed into two parts. After that, we combine them to form two new objects. Second, the two new objects are reconstructed separately using the SAKE algorithm. Finally, complete T2 data is synthesized by extracting T2 parts from each object.

#### B. K-space Synthesis for Partitioned T1/T2

Given an under-sampled T2 k-space data, our goal is to reconstruct a high-resolution T2WI. The current research has two main directions. One approach is to directly reconstruct the T2WI using a reconstruction algorithm with the aid of zero-filled T2WI. This method requires high algorithm capabilities and faces difficulties in improving the resolution of reconstructed images. Another approach is to use fully-sampled T1WI to assist under-sampled T2WI reconstruction, which is mainly conducted in image domain without feature fusion [39]. Our PKS method is distinct from previous approaches as we not only utilize fully-sampled T1 k-space data but also reconstruct the T2WI based on partial k-space synthesis. It results in more efficient and comprehensive results. Based on the above idea, we first propose two different partition mechanisms to construct different k-space data. One kind is row-based partition and the other kind is column-based partition. Row-based partition mechanism is a method of partition row by row that intercepting the k-space data in specific proportion. Similarly, the column-based mechanism is a method of partition column by column.

Since MRI sequences are obtained using a progressive scan pattern (row by row), making row-based partition more coincident with this sequential order. Thus, we adopt a row-based partition approach for the k-space data. Extensive experiments have further demonstrated the superiority of row-based partition in terms of reconstruction quality compared to column-based partition. Therefore, all subsequent descriptions are based on the execution using the row-based partition-2 mechanism. The effectiveness of row-based partition-2 method has been validated through ablation experiments. We will not elaborate on the principles of the column-based partition method, as it is the same as the row-based partition-2 method. Fig. 2 illustrates the detailed row-based partition mechanism. The method uses a 1/2 or 1/3 partition ratio. When dealing with row-based partition k-space data, the fully-sampled T1 and under-sampled T2 k-space data are proportionally divided, typically by 1/2 (Marked in green in Fig. 2). These two new k-space objects can then serve as input data in the reconstruction process. The following steps provide specific k-space partition and synthesis execution details.

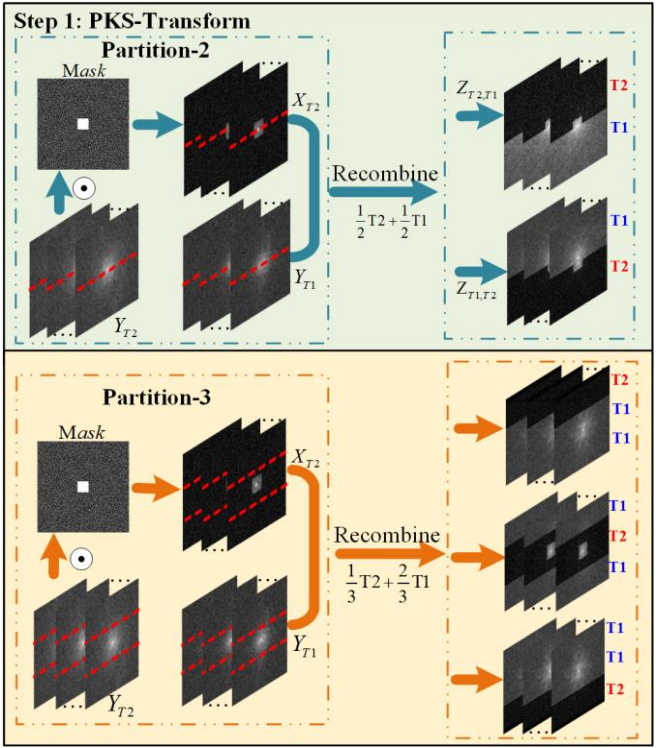

**Fig. 2.** Overview of the proposed partition-based k-space synthesis model. The green arrow and dotted boxes represent row-based partition-2 mechanism. The orange arrow and dotted boxes represent row-based partition-3 mechanism.

*Step 1: PKS-T.* To achieve high-quality T2 reconstruction, the reconstruction task of the entire under-sampled T2 k-space data is divided into two sub-problems in parallel manner. All operations performed in k-space. First, we need to obtain the objective under-sampled T2 k-space data.

$$X_{T2} = Mask \odot Y_{T2} \qquad (7)$$

Second, we decompose fully-sampled T1 k-space data and under-sampled T2 k-space data from the same slice into two in half rows respectively. In this way, the reconstruction problem of the whole T2 k-space data is transformed into two separate reconstruction problems for two equally-sized sub-T2 k-space data, which denotes as $X_{T2}(0:n/2)$, $X_{T2}(n/2+1:n)$ (represents the upper/lower half of the T2 data in the row-based partition mechanism, respectively). Similarly, sub-T1 k-space data is denoted as $Y_{T1}(0:n/2)$, $Y_{T1}(n/2+1:n)$.

$$Y_{T1} = Concat(Y_{T1}(0:n/2), Y_{T1}(n/2+1:n))^{\mathrm{T}} \qquad (8)$$
$$X_{T2} = Concat(X_{T2}(0:n/2), X_{T2}(n/2+1:n))^{\mathrm{T}} \qquad (9)$$

where $Y_{T1}$ ( $Y_{T1} = Fy_{T1}$ ) is a matrix of $n \times n$ that denotes T1 k-space data for the same slice.

Third, we combine the four-part k-space data segmented by the upper and lower halves of T1/T2 in pairs. Then assemble them into two new k-space data containing both T1 and T2 information through a concatenation operation.

$$Z_{T2,T1} = Concat(X_{T2}(0:n/2), Y_{T1}(n/2+1:n))^{\mathrm{T}} \qquad (10)$$
$$Z_{T1,T2} = Concat(Y_{T1}(0:n/2), X_{T2}(n/2+1:n))^{\mathrm{T}} \qquad (11)$$

where $Z_{T2,T1}$ and $Z_{T1,T2}$ are the synthesized new k-space data. $Z_{T2,T1}$ is the concatenation of the upper half of the k-space data of the under-sampled T2 and the lower half of the k-space data of the fully-sampled T1, and $Z_{T1,T2}$ is the concatenation of the upper half of the k-space data of the fully-sampled T1 and the lower half of the fully-sampled T2.

*Step 2: PKS-P.* In the reconstruction stage, the acquired multi-contrast k-space data is used as input for SAKE reconstruction (see Fig. 3). Detailed reconstruction implementations are provided as follows: First, the constructed two fused k-space data $Z_{T2,T1}$ and $Z_{T1,T2}$ (obtained through Step 1 transform) are sent into the SAKE separately. Second, the multi-contrast data is concatenated and represented as a Hankel matrix form (Eq. 4). Therefore, the PI reconstruction problem without calibration region can be viewed as a low-rank matrix completion problem in k-space (Eq. 6). Finally, a solution is obtained through singular value thresholding. Through the reconstruction process described above, we can obtain the reconstructed k-space data $Y_{T2,T1}$ and $Y_{T1,T2}$. For the obtained k-space data, they can be regarded as a form composed of two subparts.

$$Y_{T2,T1} = Concat(Y_{T2,T1}(0:n/2), Y_{T2,T1}(n/2+1:n))^{\mathrm{T}} \qquad (12)$$
$$Y_{T1,T2} = Concat(Y_{T1,T2}(0:n/2), Y_{T1,T2}(n/2+1:n))^{\mathrm{T}} \qquad (13)$$

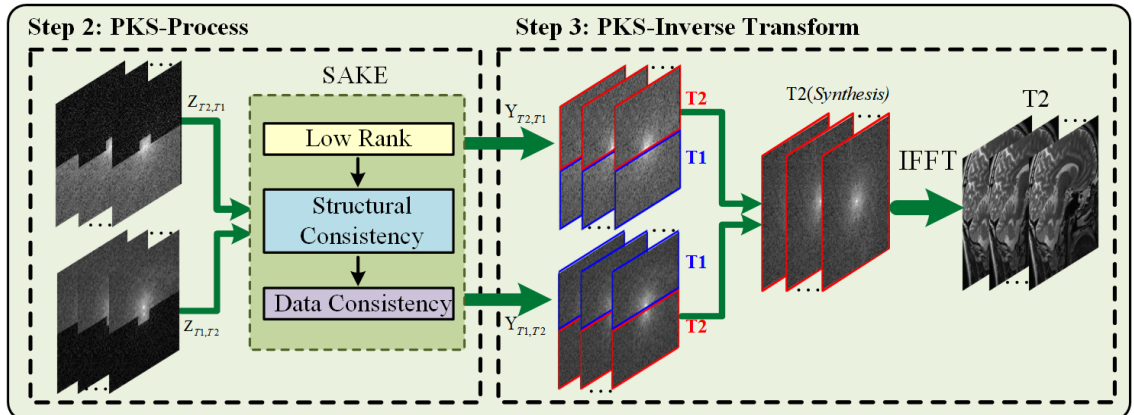

**Fig. 3.** Overview of the proposed method. Two multi-contrast k-space data of the same size as input are sent into the SAKE algorithm. Then extract the respective sub-T2 to synthesis a complete T2.

***Step 3: PKS-IT.*** After reconstruction process, the fused T2 features of the two results are aggregated together to obtain T2 representations with more comprehensive information. To achieve this, we synthesis the features belonging to the T2 part of the two reconstruction results, so that the effective information contained in the T1-assisted reconstruction can be fully propagated to the T2 to obtain the final recovery image $T2_{Syn}$, i.e.,

$$T2_{Syn} = Concat(Y_{T2,T1}(0:n/2), Y_{T1,T2}(n/2+1:n))^T \quad (14)$$

Finally, the reconstructed T2WI is obtained by performing the inverse Fourier transform of the T2 k-space data. **Algorithm 1** explains the reconstruction algorithm in detail.

---

**Algorithm 1: PKS-2**

**Transform stage**

**Input:** k-space data $Y_{T1}$, $Y_{T2}$

**1. Under sampling T2 data:** $X_{T2} = Mask \odot Y_{T2}$

**2. Decompose initial T1 data:**
$$Y_{T1}(0:n/2), Y_{T1}(n/2+1:n)$$

**3. Decompose objective T2 data:**
$$X_{T2}(0:n/2), X_{T2}(n/2+1:n)$$

**4. Recombine the first sub-T2/T1:**
$$Z_{T2,T1} = Concat(X_{T2}(0:n/2), Y_{T1}(n/2+1:n))^T$$

**5. Recombine the second sub-T1/T2:**
$$Z_{T1,T2} = Concat(Y_{T1}(0:n/2), X_{T2}(n/2+1:n))^T$$

**6. Output:** $Z_{T2,T1}$, $Z_{T1,T2}$

**Processing stage**

**1. Input:** acquired new k-space data: $Z_{T2,T1}$, $Z_{T1,T2}$

**2. Reconstruction:**
$$SAKE(LR \Rightarrow SC \Rightarrow DC)$$

**3. Output:** reconstructed k-space data:
$$Y_{T2,T1} = Concat(Y_{T2,T1}(0:n/2), Y_{T2,T1}(n/2+1:n))^T$$
$$Y_{T1,T2} = Concat(Y_{T1,T2}(0:n/2), Y_{T1,T2}(n/2+1:n))^T$$

**Inverse transform stage**

**1. Input:** $Y_{T2,T1}$, $Y_{T1,T2}$

**2. Synthesis target:**
$$T2_{Syn} = Concat(Y_{T2,T1}(0:n/2), Y_{T1,T2}(n/2+1:n))^T$$

**3. Output:** T2 k-space data

---

### C. Different Partition Strategies in PKS

In addition to the row-based partition-2 method described in **Section III. B**, we further explore the different partition methods in the PKS. We are no longer confine to a situation where the partition-2 is strictly divided equally in half. The ratio between T1 and T2 can be adjusted according to specific needs. The specific operations are detailed in Fig. 4. At the first stage, we replace the fully-sampled T1 k-space data (emphasizing on the central rows) with T2 k-space data. The number of replaced rows can be denoted by $k$ and $m$, respectively. After reconstruction, we can obtain the combined data of the reconstructed T2 and the original fully-sampled T1 k-space data. At the next stage, we need to extract the T2 portion contained in the two objects. Since the proportion of T2 k-space data is several rows more than T1 k-space data in each object, we need to average the overlapping T2 rows between the two objects before extracting T2. In his way, we can acquire three sub-blocks of T2 k-space data. At last, the target T2 is obtained by synthesizing these three sub-blocks.

The novelty of this partition strategy is that it replaces the corresponding T1 k-space data with T2. T2 k-space data can be selected according to different situations. Non-cartesian sampling is not uniform, so the number of extra T2 rows should be under-sampled. In contrast, Cartesian samples are uniformly sampled at regular intervals. When the middle low-frequency information is removed, the equivalent of the middle ACS region is fully-sampled. In this way, the number of extra T2 rows can be fully-sampled. Obviously, this partition method is more suitable for image reconstruction by Cartesian sampling method. However, the main advantage of SAKE operator is to deal with image reconstruction in non-Cartesian sampling. Therefore, to obtain the additional T2 data, we will choose the under-sampled T2 k-space data.

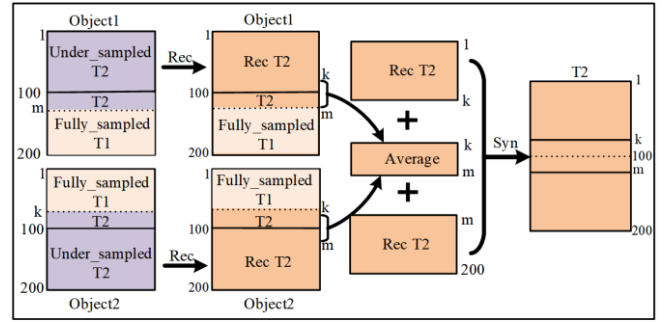

**Fig. 4.** Overview of the PKS method with different partitions.

### D. Comparison with Image-based Methods

As mentioned above, previous studies on multi-contrast images are reconstructed in image domain. The original image is directly processed and ignores the characteristics of multi-contrast images in k-space. Therefore, most methods only add auxiliary mode as prior information about the target mode, not a k-space synthesis of multi-contrast [40-43]. However, these methods are all based on deep learning networks. In terms of reconstruction results, our method cannot compare with them, thus we do not conduct direct comparative experiments. Nonetheless, from the perspective of time cost, traditional imaging methods do not require a large amount of data and training time, so it is still necessary to introduce the differences between our method and them. As shown in Fig. 5 (b). Firstly, the under-sampled T2WI is obtained as the original image. Secondly, with the help of the fully-sampled T1WI as auxiliary information, the original image is fused together with the fully-sampled T1WI and then select an appropriate reconstruction network to obtain the target T2WI.

Performing reconstruction in image domain requires computationally intensive operations such as convolution, leading to high computational complexity. On the contrary, our proposed method is unique to the previous image-based multi-contrast reconstruction. On one hand, our method involves data processing in the k-space domain. The generation of multi-contrast images can be achieved with a lower time cost, which significantly speeds up the reconstruction process and enables better handling of artifact issues. On the other hand, different from the processing methods of multi-contrast data in image domain which T1-assisted T2 imaging is a whole-based image fusion process, our method adopts the idea of partition that effectively improves the fusion ability and robustness of the model. The details are comprehensively addressed in **Section III**.

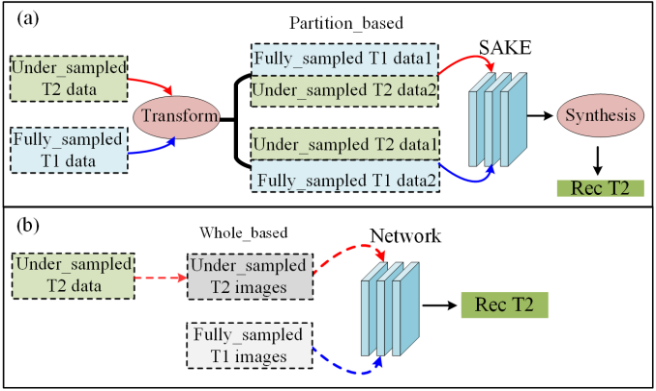

**Fig. 5.** Overview of comparison with image-based methods. (a) Partition-based k-space synthesis method. (b) Image-based synthesis method.

## IV. EXPERIMENTS

### A. Experimental Setup

In this section, we demonstrate the predominate of proposed PKS method in traditional PI algorithm. The applicability of the proposed data model under different sampling masks is tested. The generalization ability and robustness of the proposed model are evaluated. For the sake of fairness, all comparative experiments are conducted using the same parameters and computational methods. The source code is located: *https://github.com/yqx7150/PKS*.

TABLE I
MULTI-CONTRAST MRI SCAN PARAMETERS.

| Parameter | T1 | T2 | PD |
|---|---|---|---|
| TE | 11ms | 76ms | 11ms |
| TR | 700ms | 5000ms | 5000ms |
| FOV | 210mm | 210mm | 210mm |
| MATRIX SIZE | 256×256 | 256×256 | 256×256 |

*Data Acquisition:* The multi-contrast brain dataset (Fig. 6) is provided by Shenzhen Institute of Advanced Technology (**SIAT**), Chinese Academy of Sciences and informed consent of the imaging subject is obtained in accordance with Institutional Review Board policy. These are the fully-sampled k-space data for PDWI, T1WI and T2WI. The paired PDWI, T1WI and T2WI for each patient are acquired with a 3.0 T scanner. These fully-sampled data are acquired by a 12-channel head coil with matrix size of $256 \times 256$. For the sake of easier data processing, it has been truncated to $200 \times 200$. The original scan parameters for each image are shown in Table I.

*Parameter Configuration:* An iterative optimization approach is used to estimate missing k-space data based on acquired data [14]. The sliding window size is $6 \times 6$, and the number of iterations is 30. Our algorithm and the different comparison algorithms all use MATLAB, which is run on the 2 NVIDIA TITAN GPUs, 12 GB operating platform.

*Performance Evaluation:* Considering that we have pairs PDWI, T1WI and T2WI for each subject in this dataset, we can directly compare the synthesized and target mode (T2) for evaluation purposes. To objectively reflect the error caused by our presented method, we use the peak signal-to-noise ratio (PSNR) and structural similarity (SSIM) to evaluate the quality of the reconstructed images. Note that the quantitative assessments are all calculated on image domain.

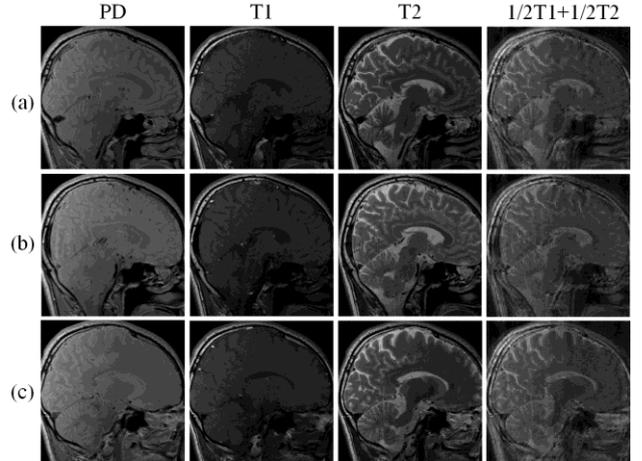

**Fig. 6.** Visualization of T1, T2 dataset and synthesized 1/2T1 and1/2 T2. 12-channel T1, T2 brain dataset form *SIAT.*

### B. Reconstruction Comparisons

For the reconstruction task, we conduct experiments on T2WI in 2D random sampling mask with acceleration factor $R = 3, 4$, 1D Cartesian sampling mask with acceleration factor $R = 2, 3$ and 2D Poisson sampling mask with acceleration factor $R = 4, 6$, respectively. In the comparison experiment section, we provide detailed descriptions of the row/column-based partition-2 method, different contrast-assisted methods, and diverse ratios synthesis of T1 to T2 for comparative analysis.

*Test on SAKE_PKS.* We compare the results of our method with those of the zero-filled reconstruction and SAKE [13] reconstruction. From Table II, we can see that the results of the SAKE algorithm are inferior to SAKE_PKS due to the lack of auxiliary information. In the 2D Random sampling mask ($R = 3$), the PSNR of row SAKE_PKS improves from 24.07 dB to 24.74 dB compared to SAKE. The effect of reconstruction quality can be determined by observing the clarity of structures in the blue error map. The more prominent the structures appear in the blue error map, the poorer the restoration outcome. Intuitively, Fig. 7 depicts the reconstruction results of the k-space data acquired by using 2D Poisson sampling masks ($R = 4$). It can be seen from the figure that the reconstructed image using the zero-filled method loses a lot of structural information. Although the SAKE reconstruction can effectively improve this problem, the reconstruction result still loses part of the details. In contrast, our approach significantly reduces aliasing and noise while better preserving edges and structural information in the reconstructed image. This enhances the level of detail, reduces observational errors, and provides clearer image contour information.

TABLE II
PSNR AND SSIM COMPARISON WITH TRADITIONAL PI METHOD UNDER 2D RANDOM AND POISSON PATTERNS WITH VARYING ACCELERATION FACTORS.

| *(a)* | Zero-filled | SAKE | Column SAKE_PKS | Row SAKE_PKS |
|---|---|---|---|---|
| 2D Random *R*=3 | 20.16/0.6220 | 24.07/0.7772 | **24.67/0.7853** | **24.74/0.7871** |
| 2D Random *R*=4 | 19.29/0.5578 | 22.09/0.7017 | **22.60/0.7097** | **22.62/0.7103** |
| 2D Poisson *R*=4 | 21.35/0.6704 | 25.94/0.8171 | **26.38/0.8216** | **26.48/0.8224** |
| 2D Poisson *R*=6 | 20.46/0.6139 | 24.02/0.7620 | **24.52/0.7695** | **24.59/0.7719** |

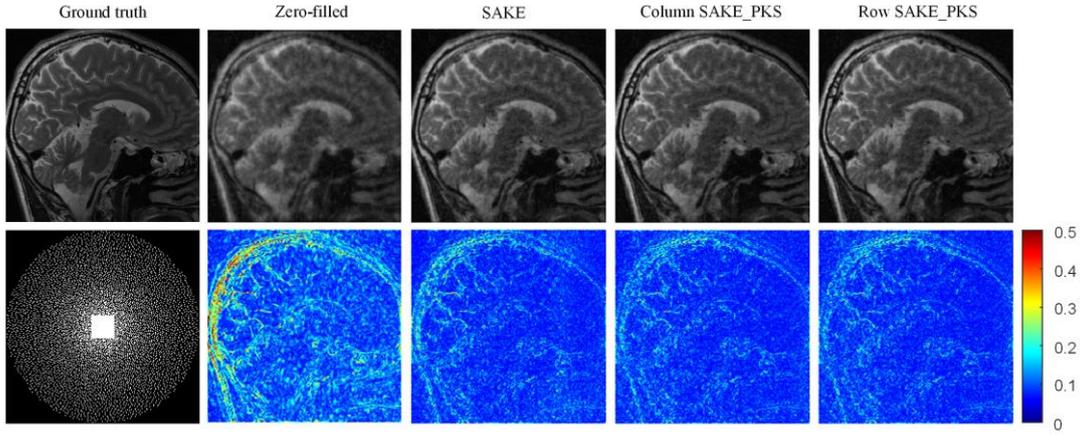

**Fig. 7.** Complex-valued PI reconstruction results at *R*=6 using Poisson mask with 12 coils (Fig. 5 (a)). From left to right: Ground truth, Zero-filled, reconstruction by SAKE, Column SAKE_PKS and Row SAKE_PKS.

*Test on Different Multi-contrast.* The first experiment demonstrates the effectiveness of the PKS method in T1 assisting T2. Building on these findings, we find that is also feasible to further enhance T2 reconstruction by utilizing other contrast such as PD. Additionally, another approach involves leveraging the combination of T1 and PD to provide supplementary assistance for T2 reconstruction. According to the results from Table III, it can be observed that whether it is T1 assisting T2, PD assisting T2, or both T1 and PD jointly assisting T2, the effect of assisted T2 reconstruction is better compared to independent T2 reconstruction. Particularly, the best reconstruction effect for T2 is achieved when T1 and PD jointly assisting T2. Especially under 1D Cartesian $R = 2$, the PSNR for T1 and PD jointly assisting T2 increases by about 1.21 dB compared to T2 alone. This suggests that combining information from multi-contrast MR images can significantly improve the quality of reconstruction in some cases. Furthermore, from the reconstructed images in Fig. 8, the residuals reconstructed for SAKE have more severe overlap artifacts. In contrast, there are significantly fewer overlapping artifacts in image reconstruction using multi-contrast synthesized images. The reconstruction error of our method is smaller and the closest to the target image, which enables a more accurate reconstruction.

TABLE III
PSNR AND SSIM COMPARISON WITH TRADITIONAL PI METHOD UNDER 2D RANDOM AND 1D CARTESIAN PATTERNS WITH VARYING ACCELERATION FACTORS.

| (b) | Zero-filled | T2 | 1/2T1+1/2T2 | 1/2PD+1/2T2 | 1/4T1+1/4PD+1/2T2 |
|---|---|---|---|---|---|
| 1D Cartesian *R*=2 | 22.71/0.7421 | 27.85/0.8699 | 28.76/0.8766 | 29.03/0.8784 | **29.06/0.8786** |
| 1D Cartesian *R*=3 | 21.33/0.6802 | 24.85/0.8060 | 25.53/0.8106 | 25.70/0.8108 | **25.72/0.8109** |
| 2D Random *R*=3 | 20.86/0.6195 | 24.70/0.7727 | 25.31/0.7806 | 25.49/0.7825 | **25.51/0.7829** |
| 2D Random *R*=4 | 20.05/0.5613 | 22.78/0.7003 | 23.34/0.7107 | 23.45/0.7114 | **23.46/0.7166** |

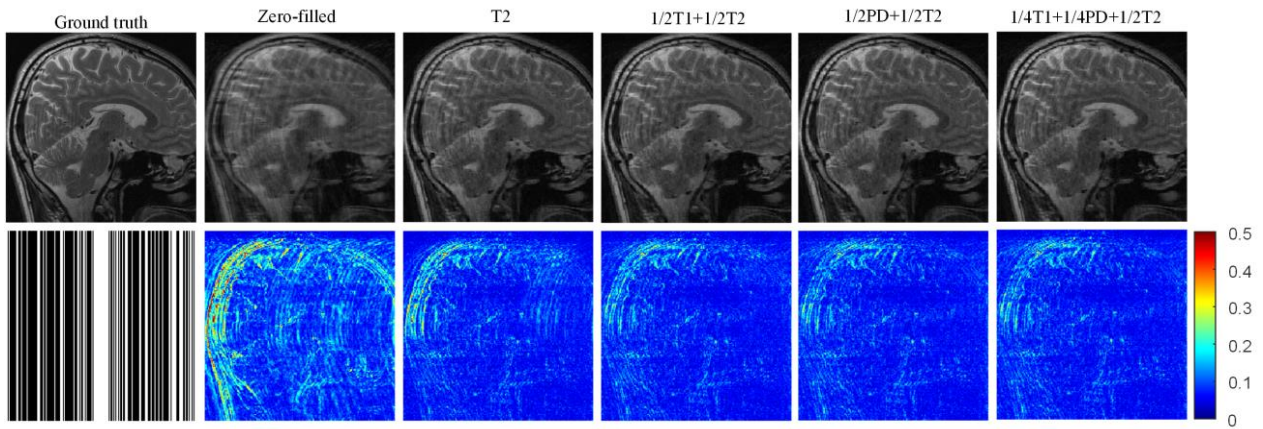

**Fig. 8.** Complex-valued PI reconstruction results at *R*=3 using 1D Cartesian mask with 12 coils (Fig. 5 (b)). From left to right: Ground truth, Zero-filled, reconstruction by T2, T1 assisting T2, PD assisting T2 and T1 and T2 jointly assisting T2.

*Test on Diverse Ratios Synthesis of T1 to T2.* In previous experiments, we use different partition mechanisms for T1 assisting T2, and try to test the PKS method for T2 reconstruction with different contrast. All these methods have achieved good results. However, to further investigate the effect of different ratios of auxiliary mode and target mode on the PKS method, we conduct further tests. In the above PKS method, the auxiliary mode and the target mode are divided in equal proportion, that is, the data of T1 and T2 are combined in a ratio of 1/2 (1/2T1+1/2T2). To explore the effects of different ratios, we try different ratios of auxiliary mode and target mode. As expected, the results from Table IV indicate that when the T2 k-space data is 5 rows more than T1 (+5), although the effect of equal proportion segmentation can be achieved, the reconstruction effect gradually deteriorates with the increase of the rows of T2 k-space data. Surprisingly, this approach is still able to produce reconstruction results comparable to or even better than T2. It can also be observed from Fig. 9 that the residual plot of the half partition and the T2 k-space data is 5 rows more than T1 (+5) show less prominent edges, indicating clearer reconstructed images.

Overall, our model outperforms SAKE on all above PKS methods. This demonstrates the robustness of the proposed PKS and indicates that it can effectively reconstruct target images.

TABLE IV
PSNR AND SSIM COMPARISON WITH TRADITIONAL PI METHOD UNDER 2D RANDOM AND POISSON PATTERNS WITH VARYING ACCELERATION FACTORS.

| (c) | Zero-filled | T2 | 1/2T1+1/2T2 | 1/2T1+1/2T2(+5) | 1/2T1+1/2T2(+10) |
|---|---|---|---|---|---|
| 2D Random $R$=3 | 20.29/0.6281 | 24.39/0.7672 | **25.05/0.7735** | **25.07/0.7751** | 24.93/0.7731 |
| 2D Random $R$=4 | 19.33/0.5659 | 22.31/0.6978 | **22.77/0.7015** | **22.80/0.7043** | 22.76/0.7038 |
| 2D Poisson $R$=4 | 21.48/0.6693 | 26.04/0.8031 | **26.54/0.8074** | **26.57/0.8084** | 26.52/0.8080 |
| 2D Poisson $R$=6 | 20.31/0.6056 | 23.90/0.7429 | **24.49/0.7496** | **24.51/0.7503** | 24.41/0.7489 |

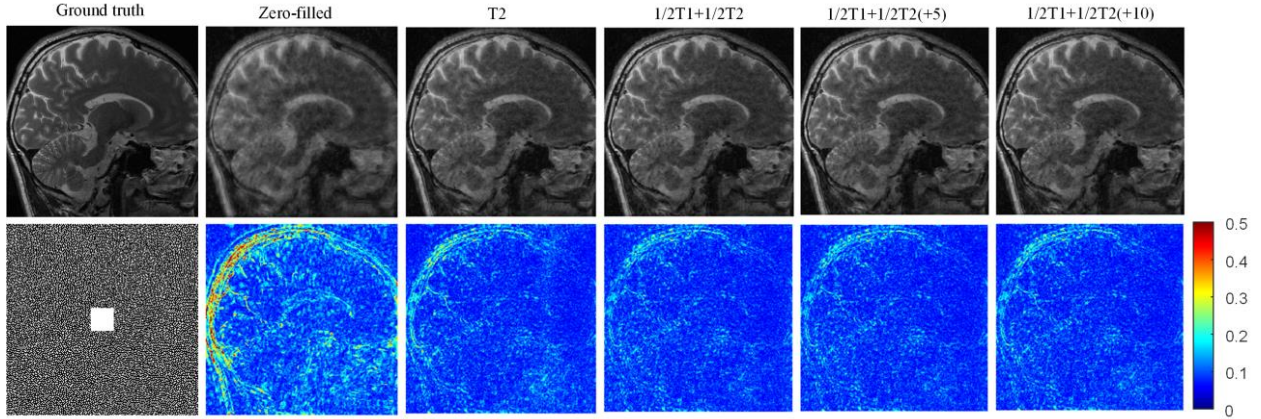

Fig. 9. Complex-valued PI reconstruction results at $R$=3 using Random mask with 12 coils (Fig. 5 (c)). From left to right: Ground truth, Zero-filled, reconstruction by T2, equal ratios of T1 to T2, 5 more rows of T2 k-space data (+5) and 10 more rows of T2 k-space data (+10).

## C. Computational Cost

To evaluate the performance of the proposed method, we compare the reconstruction time for each approach. Table V records the reconstruction times for each partition method. It can be observed that as the number of partition increases, the reconstruction time for the PKS method also increases. On one hand, this is because the PKS method utilizes multiple contrast data for reconstruction, which inherently requires more time compared to single contrast reconstruction. On the other hand, our approach transforms the recovery problem of one data into multiple data, resulting in longer reconstruction times with increasing partition numbers. However, when considering the average reconstruction time per partition, our partition methods are similar and comparable to SAKE algorithm. More importantly, the generation of multi-contrast images can be achieved with a lower time cost than deep learning methods, which significantly speeds up the reconstruction process. In the future research, we will explore more suitable algorithms to reduce the reconstruction time of the PKS method and further enhance its efficiency.

TABLE V
THE RECONSTRUCTION TIME REQUIRED FOR EACH METHOD.

| (b) | Total-time (s) | Single-time (s) |
|---|---|---|
| SAKE | 56.03 | 56.03 |
| Partition-2 | 127.07 | 63.55 |
| Partition-3 | 192.76 | 64.25 |
| Partition-4 | 255.03 | 63.76 |

## D. Ablation Study

We validate the performance of the proposed PKS method by employing various forms of partition methods. In this experiment, we decompose the k-space data using different block partition techniques. The quantitative results of the experiment are presented in Table VI, illustrating the comparison of different partition methods in terms of reconstruction performance. When selecting Random sampling mask $R$ = 3, 4, it is evident that regardless of using partition-2, partition-3, or partition-4 for block partition, the row-based partition consistently outperforms the column-based partition. The underlying reason behind this phenomenon is that magnetic resonance sequence scanning is conducted row by row, making row-based partition more aligned with this sequential order. As a result, row-based partition can better leverage the information in the dataset, leading to superior reconstruction results.

TABLE VI
PSNR AND SSIM COMPARISON WITH TRADITIONAL PI METHOD UNDER RANDOM SAMPLING PATTERNS WITH VARYING ACCELERATION FACTORS.

| Partition-2 | 2D Random $R$=3 | 2D Random $R$=4 |
|---|---|---|
| Column SAKE_PKS | 25.19/0.7775 | 23.18/0.7038 |
| Row SAKE_PKS | **25.31/0.7806** | **23.34/0.7107** |
| Partition-3 | 2D Random $R$=3 | 2D Random $R$=4 |
| Column SAKE_PKS | **25.20/0.7798** | **23.20**/0.7076 |
| Row SAKE_PKS | 25.09/0.7794 | 23.14/**0.7086** |
| Partition-4 | 2D Random $R$=3 | 2D Random $R$=4 |
| Column SAKE_PKS | 25.35/0.7797 | 23.24/0.7048 |
| Row SAKE_PKS | **25.53/0.7855** | **23.46/0.7130** |

To further verify the impact of the number of various blocks under the same partition mechanism, we use SAKE_PKS method to explain these phenomena. It can be concluded from Table VII that under Poisson 4 and 6 sampling, each of the three methods has its advantages and disadvantages. When based on row partition mechanism, parti-

tion-4 demonstrates the best performance, whereas based on column partition mechanism, partition-3 performs the best. At the same time, the performance of partition-2 both remains at a moderate level. However, due to the utilization of a traditional algorithm, achieving a partition-3 or partition-4 requires a longer processing time. Taking all factors into consideration, we have determined that partition-2 is the most appropriate choice.

TABLE VII
PSNR AND SSIM COMPARISON WITH TRADITIONAL PI METHOD UNDER POISSON SAMPLING PATTERNS WITH VARYING ACCELERATION FACTORS.

| Column SAKE_PKS | 2D Poisson $R$=4 | 2D Poisson $R$=6 |
|---|---|---|
| **Partition-2** | 26.89/0.8147 | 24.89/0.7559 |
| **Partition-3** | **27.04/0.8173** | **25.02/0.7602** |
| **Partition-4** | 26.88/0.8129 | 24.94/0.7561 |
| **Row SAKE_PKS** | 2D Poisson $R$=4 | 2D Poisson $R$=6 |
| **Partition-2** | 27.13/0.8204 | 25.07/0.7621 |
| **Partition-3** | 27.11/0.8200 | 24.99/0.7632 |
| **Partition-4** | **27.19/0.8209** | **25.17/0.7640** |

After verifying the effects of row-column partition and different partition numbers under same partition mechanism, we further investigate the impact of using T1 and T2 in different proportions under the context of row-based partition-2. Frankly speaking, there is a tendency to play a very important role of auxiliary modality images (T1) in multi-contrast images. The result shown in Table VII has explained this phenomenon from different study subjects (partition-2, 3, 4). Subsequently, we examine the reconstruction effect of T1-assisted T2 with incomplete equal proportions from the same study subject using the row-based partition-2 method. It can be clearly inferred from Table VIII that, although it is possible to achieve the effect of equal proportion partition when the T2 k-space data is 5 rows more than T1 (+5), the reconstruction effect gradually deteriorates with the increasing amount of T2 k-space data. On one hand, auxiliary modality images typically possess richer contrast and specific tissue information, which can provide additional anatomical structures. Consequently, reducing the auxiliary mode component leads to a decrease in the accuracy of reconstruction. On the other hand, non-cartesian sampling is not uniform, so the number of extra T2 rows should be under-sampled. Therefore, we replace the fully-sampled T1 k-space data with under-sampled T2, the quality of the reconstruction will gradually decline with the increasing substitution of under-sampled T2.

TABLE VIII
PSNR AND SSIM COMPARISON WITH TRADITIONAL PI METHOD UNDER POISSON SAMPLING PATTERNS WITH VARYING ACCELERATION FACTORS.

| Row SAKE_PKS | 2D Poisson $R$=4 | 2D Poisson $R$=6 |
|---|---|---|
| 1/2T1+1/2T2 | 27.12/0.8204 | **25.07**/0.7621 |
| 1/2T1+1/2T2(+5) | **27.14/0.8216** | 25.06/**0.7636** |
| 1/2T1+1/2T2(+10) | 27.10/0.8214 | 24.99/0.7629 |
| 1/2T1+1/2T2(+15) | 27.02/0.8203 | 24.90/0.7619 |
| 1/2T1+1/2T2(+20) | 26.92/0.8189 | 24.84/0.7613 |

## V. DISCUSSION

This work presented a novel PKS multi-contrast method. The motivation behind our PKS approach stem from the fact that multi-contrast imaging has longer acquisition time and is easy to cause motion artifacts. Existing multi-contrast methods solely rely on incorporating auxiliary modes as prior information for the target mode to enhance image quality. However, they utilize whole-based image domain fusion methods or do not perform partial k-space synthesis between different contrasts. These limitations result in intensive computational workload and poor flexibility. To alleviate these issues, we further explored this aspect by utilizing fully-sampled T1 k-space data. We then used partial k-space synthesis and combined it with traditional PI technology SAKE to reconstruct the T2WI. This methodology produced more efficient and comprehensive reconstruction results.

The experiment demonstrated that the proposed PKS method greatly attenuates aliasing and noise. At the same time, it reliably recovered fine details and suppressed artifacts, thus leading to more comprehensive and interpretable reconstructions. In the comparison experiment, we compared our PKS method with traditional PI algorithm SAKE. Unlike SAKE that processes each contrast independently, SAKE_PKS adopts multiple contrasts to assist T2 reconstruction. All of them achieved superior image quality compared to the single contrast. More importantly, this method not only introduced various partition mechanisms, but also used different auxiliary modes and different proportions of multi-contrast synthesis. The superior performances verified the robustness of the PKS model. Furthermore, k-space fusion process significantly speeded up the reconstruction process and enabled better handling of artifact issues.

Although the proposed PKS is superior to some traditional PI methods both subjectively and objectively, there are still some limitations that need to be tackled. The first option is the problem existing in the traditional algorithm itself, because our algorithm does not change the inside of the traditional algorithm, but only changes their input, so it needs to improve the algorithm optimization to further improve the quality of the reconstructed image. Furthermore, because our method adopts the idea of partition, each part needs to be reconstructed separately, so the application time will be longer, which is also the biggest shortcoming of our method.

## VI. CONCLUSION

This work proposed a PKS multi-contrast method to assist in target mode image reconstruction. The model can accelerate the reconstruction of under-sampled T2 with the assistance of fully-sampled auxiliary mode. Unlike existing multi-contrast reconstruction, the model performs all operations in k-space and achieves asymptotic feature fusion. We have conducted extensive experiments on different multi-contrast, diverse ratios of T1 to T2 and different sampling masks to demonstrate the generalization and robustness of our proposed model. Experimental results illustrated that the developed PKS scheme can successfully reconstruct the under-sampled target modal image, and had excellent performance in restoring image quality and preserving details. At the same time, our method showed good adaptability and robustness under different contrast-assisted and T1-T2 ratios. Efficient target modal image reconstruction under various conditions were realized. In future study, we will broaden our model's capabilities by incorporating state-of-the-art algorithms and exploring diverse partition methods, such as alternating n-times for multiple iterations. These efforts aim to validate the model's flexibility and enhance its overall performance.

## REFERENCE

[1] A. J. Sederman, L. F. Gladden, and M. D. Mantle, "Application of magnetic resonance imaging techniques to particulate systems," *Adv. Powder. Techn.*, vol. 18, pp. 23–38, 2007.

[2] R. Stannarius, "Magnetic resonance imaging of granular materials," *Rev. Sci. Instr.*, vol. 88, no. 5, 2017.

[3] S. D. Serai. "Basics of magnetic resonance imaging and quantitative parameters T1, T2, T2*, T1rho and diffusion-weighted imaging," *Pediatric radiology.*, vol. 52, no. 2, pp. 217-227, 2022.

[4] P. Maillard, N. Delcroix, F. Crivello *et al.* "An automated procedure for the assessment of white matter hyperintensities by multispectral (T1, T2, PD) MRI and an evaluation of its between-centre reproducibility based on two large community databases," *Neuroradiology*, vol.50, pp. 31–42, 2008.

[5] Z.-P. Liang and P. C. Lauterbur, "Principles of magnetic resonance imaging: a signal processing perspective", *SPIE Optical Engineering Press*, 2000.

[6] D. W. McRobbie, E. A. Moore et al., "MRI from Picture to Proton", *Cambridge university press*, 2007.

[7] S. Ravishankar, Y. Bresler. "MR image reconstruction from highly under-sampled k-space data by dictionary learning", *IEEE Trans. Med. Imag.*, vol. 30, no. 5, pp. 1028-1041, 2010.

[8] C. M. Tsai, D. G. Nishimura, "Reduced aliasing artifacts using variable-density k-space sampling trajectories", *Magn. Reson. Med.*, vol.43, no. 3, pp. 452-458, 2000.

[9] K. P. Pruessmann, M. Weiger, M. B. Scheidegger, and P. Boesiger, "SENSE: Sensitivity encoding for fast MRI," *Magn. Reson. Med.*, vol. 42, no. 5, pp. 952-962, 1999.

[10] M. A. Griswold, P. M. Jakob, R. M. Heidemann, M. Nittka, V. Jellus, J. Wang, B. Kiefer, and A. Haase, "Generalized autocalibrating partially parallel acquisitions (GRAPPA)," *Magn. Reson. Med.*, vol. 47, pp. 1202–1210, 2002.

[11] M. Lustig, J. M. Pauly, "SPIRiT: iterative self-consistent parallel imaging reconstruction from arbitrary k-space", *Magn. Reson. Med.*, vol. 64, no. 2, pp. 457-471,2010.

[12] M. Uecker, P. Lai, M. J. Murphy, et al., "ESPIRiT—an eigenvalue approach to autocalibrating parallel MRI: where SENSE meets GRAPPA," *Magn. Reason. Med.*, vol. 71, no. 3, pp. 990-1001, 2014.

[13] P. J. Shin, P. E. Larson, M. A. Ohliger, et al., "Calibrationless parallel imaging reconstruction based on structured low-rank matrix completion," *Magn. Reson. Med.*, vol. 72, no. 4, pp. 959-970, 2014.

[14] J. P. Haldar, J. Zhuo, "P-LORAKS: Low-rank modeling of local k-space neighborhoods with parallel imaging data," *Magn. Reson. Med.*, vol. 75, no. 4, pp. 1499-1514, 2016.

[15] Q. Lyu, H. Shan, C. Steber, C. Helis, C. Whitlow, M, Chan, G. Wang, "Multi-contrast super-resolution MRI through a progressive network," *IEEE Trans. Med. Imag.*, vol. 39, no. 9, pp. 2738-2749, 2020.

[16] J. Huang, C. Chen, and L. Axel. "Fast multi-contrast MRI reconstruction," *Magn. Reson. Imag.*, vol. 32, no. 10, pp. 1344-1352, 2014.

[17] B. Bilgic, V. K. Goyal, and E. Adalsteinsson, "Multi-contrast reconstruction with Bayesian compressed sensing," *Magn. Reson. Med.*, vol. 66, no. 6, pp. 1601-1615, 2011.

[18] X. Liu, M. Zhang, Q. Liu, "Multi-contrast MR reconstruction with enhanced denoising autoencoder prior learning," 2020 IEEE 17th International Symposium on Biomedical Imaging (ISBI), pp. 1-5, 2020.

[19] K. H Kim., W.-J. Do, and S.-H. Park, "Improving resolution of MR images with an adversarial network incorporating images with different contrast," *Med. Phys.*, vol. 45, pp. 3120–3131, 2018.

[20] C. Alkan, J. Cocjin, A. Weitz, "Magnetic resonance contrast prediction using deep learning." *Google Scholar*, 2016.

[21] R. Vemulapalli, H. Van Nguyen, and S. Kevin Zhou, "Unsupervised crossmodal synthesis of subject-specific scans," *Proc. IEEE Int. Conf. Comput. Vis.*, pp. 630–638, 2015.

[22] L. Xiang, Y. Chen, W. Chang, "Deep-learning-based multi-modal fusion for fast MR reconstruction," *IEEE Trans. Bio. Eng.*, vol. 66, no. 7, pp. 2105-2114, 2018.

[23] V. Sevetlidis, M. V. Giuffrida, and S. A. Tsaftaris, "Whole image synthesis using a deep encoder-decoder network," in *Simulation and Synthesis in Medical Imaging*, 2016, pp. 127–137.

[24] O Dalmaz, M Yurt, T Çukur, "ResViT: Residual vision transformers for multimodal medical image synthesis," IEEE Trans. Med. Imag., vol. 41, no. 10, pp. 2598-2614, 2022.

[25] J. A. Cadzow. "Signal enhancement-a composite property mapping algorithm," *IEEE Trans. Acoustics, Speech, and Signal Processing*, vol. 36, no. 1, pp. 49-62, 1988.

[26] J. Zhang, C. Liu, M. E. Moseley, "Parallel reconstruction using null operators," Magn. Reason. Med., vol. 66, no. 5, pp. 1241-1253, 2011.

[27] M. Lustig, M. Elad, J.M. Pauly, "Calibrationless parallel imaging reconstruction by structured low-rank matrix completion," *Proc. Int. Soc. Magn. Reson. Med.*, pp. 2870, 2010.

[28] M. Lustig, "Post-cartesian calibrationless parallel imaging reconstruction by structured low-rank matrix completion," *Proc. Int. Soc. Magn. Reson. Med.*, vol. 483, 2011.

[29] A. Chartsias, T. Joyce, M. V. Giuffrida, and S. A. Tsaftaris, "Multimodal MR synthesis via modality-invariant latent representation," *IEEE Trans. Med. Imag.*, vol. 37, no. 3, pp. 803–814, 2018.

[30] T. Joyce, A. Chartsias, and S. A. Tsaftaris, "Robust multi-modal MR image synthesis," in *MICCAI*, pp. 347–355, 2017.

[31] A. Beers, J. Brown, K. Chang, J. Campbell, S. Ostmo, M. Chiang, andJ. Kalpathy-Cramer, "High-resolution medical image synthesis using progressively grown generative adversarial networks," *arXiv:1805.03144*, 2018.

[32] T. Zhou, H. Fu, G. Chen, J. Shen, and L. Shao, "Hi-net: Hybrid-fusion network for multi-modal MR image synthesis," *IEEE Trans. Med. Imag.*, vol. 39, no. 9, pp. 2772–2781, 2020.

[33] A. Sharma and G. Hamarneh, "Missing MRI pulse sequence synthesis using multi-modal generative adversarial network," *IEEE Trans. Med. Imag.*, vol. 39, pp. 1170–1183, 2020.

[34] B. Murugesan, S. Ramanarayanan, S. Vijayarangan, "A deep cascade of ensemble of dual domain networks with gradient-based T1 assistance and perceptual refinement for fast MRI reconstruction," *Comput. Med. Imag. Grap.*, vol. 91, 2021.

[35] Q. Yang, N. Li, Z. Zhao, "Mri cross-modality neuroimage-to-neuroimage translation," *arXiv preprint arXiv:1801.06940*, 2018.

[36] D. Kawahara, Y. Nagata, "T1-weighted and T2-weighted MRI image synthesis with convolutional generative adversarial networks,". *Rep. Pract. Oncol. Radiother.*, vol. 26, no. 1, pp. 35-42, 2021.

[37] S. U. Dar, M. Yurt, M. Shahdloo, M. E. Ildız, B. Tınaz, T. Cukur, "Prior-guided image reconstruction for accelerated multi-contrast mri via generative adversarial networks," *IEEE J-STSP*, vol. 14, no. 6, pp. 1072–1087, 2020.

[38] K. H. Kim, W.-J. Do, S.-H. Park, "Improving resolution of mr images with an adversarial network incorporating images with different contrast," *Med. Phys.*, vol. 45, no. 7, pp. 3120–3131, 2018.

[39] X. Liu, J. Wang, H. Sun, "On the regularization of feature fusion and mapping for fast MR multi-contrast imaging via iterative networks," *Magn. Reson. Imag.*, vol. 77, pp. 159-168, 2021.

[40] B. Zhou, S. K. Zhou, "Dudornet: Learning a dual-domain recurrent network for fast mri reconstruction with deep t1 prior," *Proceedings of the IEEE/CVF conference on computer vision and pattern recognition*, pp. 4273–4282, 2020.

[41] A. Bustin, G. Lima da Cruz, O. Jaubert, "High-dimensionality undersampled patch-based reconstruction (HD-PROST) for accelerated multi-contrast MRI," *Magn. Reson. Med.*, vol. 81, no. 6, pp. 3705-3719, 2019.

[42] P. Song, L. Weizman, J. Mota, "Coupled dictionary learning for multi-contrast MRI reconstruction," *IEEE Trans. Med. Imag.*, vol. 39, no. 3, pp. 621-633, 2019.

[43] V. Bhateja, M. Nigam, A. S. Bhadauria, "Two-stage multi-modal MR images fusion method based on parametric logarithmic image processing (PLIP) model," *Pattern. Recogn. Lett.*, vol. 136, pp. 25-30, 2020.